\documentstyle[aps,prl,twocolumn,subfigure,epsfig,amsmath,amssymb]{revtex}

\begin{document}
\title{Dynamics of trapped two-component Fermi gas: temperature dependence of the transition from collisionless to collisional regime}
\author{F. Toschi$^{1,2}$, P. Vignolo$^3$, S. Succi$^{1,3}$ and M.P. Tosi$^3$}
\address{
$^1$ Istituto per le Applicazioni del Calcolo, CNR, Viale del Policlinico 137, I-00161 Roma, Italy\\
$^2$ INFM, Sezione di Roma ``Tor Vergata'', Via della Ricerca Scientifica 1, I-00133 Roma, Italy\\
$^3$ NEST-INFM and Classe di Scienze, Scuola Normale Superiore,
Piazza dei Cavalieri 7, I-56126 Pisa, Italy}
\maketitle

\begin{abstract}
We develop a numerical method to study the dynamics 
of a two-component atomic Fermi gas trapped inside a
harmonic potential at temperature $T$ well below the Fermi temperature
$T_F$.
We examine the transition from the collisionless to the collisional 
regime down to $T=0.2\,T_F$ and find
good qualitative agreement with the experiments 
of B. DeMarco and D.S. Jin [Phys. Rev. Lett. {\bf 88}, 040405 (2002)].
We demonstrate a twofold role of temperature on
 the collision rate and on the efficiency of collisions. 
In particular we observe an hitherto unreported effect, namely
that the transition to hydrodynamic behavior is 
shifted towards lower collision rates as temperature decreases.
\vskip 0.2cm
\end{abstract}
\vskip0.3cm

\vskip0.2cm

Recent experiments at JILA~\cite{gensemer}
on the collisions 
between two oscillating spin-polarized components of a Fermi gas 
of $^{40}$K atoms have shown
that this setup is an important tool for investigating
the dynamics of dilute quantum gases. These experiments 
have given evidence for a
transition from collisionless (zero-sound) to hydrodynamical 
(first-sound) behavior: 
the measured damping time $\tau$ becomes very long
at both low and large values of the estimated collision rate.
The transition from the hydrodynamic to an intermediate regime 
has also been followed with decreasing temperature and the effect of 
Pauli blocking of the collisions
from increasing occupation of final states has been exhibited
at low temperature.

Various 
numerical experiments have addressed the dynamics of  
a thermal cloud of bosons~\cite{foot}, even in the presence of a 
Bose-Einstein condensate~\cite{zaremba}
or of cold spin-polarized fermions interacting with a condensate
{\it via} mean field~\cite{marilu}.
Fermion molecular dynamics (FMD) has been developed~\cite{wilet}
as a quasi-classical
model for treating problems such as ion-atom collisions or
formation of antiprotonic atoms.
However, FMD is not adequate to deal with
dilute fermionic systems, such as the neutral
atomic gas under harmonic confinement
realized in Ref.~\cite{gensemer},
and an approach explicitly acknowledging
the diluteness and other characteristics of such a system is needed.
To our knowledge, the present work reports
the first numerical study directed to  
the transport properties of ultracold Fermi gases. 

In this Letter we solve the Vlasov-Landau equations (VLE) 
for two-component fermionic Wigner distributions.
As in FMD~\cite{wilet} or in numerical studies of the dynamics of thermal 
bosons~\cite{foot,zaremba,marilu}, the quantum-mechanical fluid 
is treated by a particle-dynamics approach.
We proceed along the path traced
for a single spin-polarized Fermi gas~\cite{marilu},
by duplicating it and 
introducing mean-field interactions
and collisions between the two components.
A major highlight of our numerical
method is the strategy used to deal with collisional events. 
We develop a {\it locally adaptive importance-sampling} 
technique which allows us to handle collisional interactions 
faster than 
in standard Monte Carlo techniques
by several orders of magnitude.

Owing to this computational advance we are able to
examine collisional effects down to $T/T_F \sim 0.2$, in a 
region of temperature $T$ well below the Fermi temperature $T_F$
where Pauli blocking 
would normally grind the simulation to a halt because of
numerical attrition problems (basically through a saturation of 
phase space resulting in vanishing efficiency of the Monte Carlo 
sampling). 
Moreover, since we focus on the JILA setup~\cite{gensemer}
where the axial symmetry is maintained during the experiment,
we can use an axially symmetric code with two-dimensional collisions,
in which the angular degree of freedom is taken into account {\it via} an
effective weight.
The resulting code permits us to 
study the collisional properties of the two-component
Fermi gas as functions of both temperature $T$ and 
quantum collision rate $\Gamma_q$. 
The numerical approach has full control over
additional variables which are experimentally unaccessible:
in particular, while in the experiments the {\it classical}
collision rate can easily be estimated but 
the {\it fully quantal} collision rate $\Gamma_q$ remains unknown,
in the numerical approach both classical and
quantum collision rates can be counted step by step.
Because of this control of the collision rates we are
able to observe that, even if most
collisions become forbidden classically and by the Pauli principle
as temperature is lowered, the 
few collisions that can occur involve particles 
increasingly clustered around the Fermi level. 
The result is a kind of ``catalytic effect'' 
in which these few collisions 
suffice to drive the gas from the collisionless 
to the hydrodynamic regime.

\paragraph*{The model and its solution.}

Here we write the VLE for two interacting
Fermi gases and describe the algorithm used in their
numerical solution.
We summarize some points which have been made
in more detail in Ref.~\onlinecite{marilu} and point out
how the difficulties 
raised by Fermi statistics are handled in our method.

The two fermionic components in external
potentials $V_{ext}^{(j)}({\bf r})$ are described by the distribution functions 
$f^{(j)}({\bf r},{\bf p},t)$ with $j=1$ or 2.
These obey the kinetic equations
\begin{equation}
\partial_t f^{(j)}+\dfrac{\bf p}{m}\cdot{\bf \nabla}_{{\bf r}} 
f^{(j)}-
{\bf \nabla}_{{\bf r}} U^{(j)}\cdot {\bf \nabla}_{{\bf p}}f^{(j)}=
C_{12}[f^{(j)}]
\label{vlasov}
\end{equation}
where the mean-field Hartree-Fock (HF) effective potential is
$U^{(j)}({\bf r},t)\equiv V_{ext}^{(j)}
({\bf r})+g n^{(\overline{j})}({\bf r},t)$
with $\overline{j}$ denoting the species different from $j$. 
Here we have set $\hbar=1$, $g=2\pi a/m_r$ with
$a$ being the $s$-wave scattering length between two atoms
and $m_r$ the reduced mass, and $n^{(j)}({\bf r},t)$ is
the spatial density given by integration of 
$f^{(j)}({\bf r},{\bf p},t)$ over momentum.

Collisions between atoms of the same spin 
can be neglected at low temperature, so that in Eq. (\ref{vlasov}) the term $C_{12}$ 
involves only collisions between particles with different polarizations:
\begin{eqnarray}
&&C_{12}[f^{(j)}]\equiv
2(2\pi)^4 g^2/V^3\nonumber\\
&&\sum_{{\bf p}_2,{\bf p}_3,{\bf p}_4} 
\Delta_{\bf p}\Delta_{\varepsilon}
\,[{\overline f}^{(j)}{\overline f}_2^{(\overline{j})}f_3^{(j)}
f_4^{(\overline{j})}-
f^{(j)}f^{(\overline{j})}_2{\overline f}_3^{(j)}
{\overline f}_4^{(\overline{j})}]
\label{integralcoll}
\end{eqnarray}
with $f^{(j)}\equiv f^{(j)}({\bf r},{\bf p},t)$, 
${\overline f}^{(j)} \equiv 1-f^{(j)}$,  
$f_i^{(j)}\equiv f^{(j)}({\bf r},{\bf p}_i,t)$, 
${\overline f}_i^{(j)} \equiv 1-f_i^{(j)}$.
$V$ is the volume occupied by the gas and
the factors $\Delta_{\bf p}$ and $\Delta_{\varepsilon}$ are the usual delta
functions accounting for conservation of momentum
and energy, with the energies given by $p_i^2/2m_j+U^{(j)}({\bf r},t)$.

The numerical procedure by which the VLE is advanced in time consists of three
basic steps: ({\it i}) initialization of the fermionic distributions, 
({\it ii}) propagation in phase space, and 
({\it iii}) collisional interactions.

The initial distributions in equilibrium at the bottom of
a harmonic trap are generated by using the HF expression
\begin{equation}
f^{(j)}_{eq}({\bf r},{\bf p})=\left\{\exp\left[\beta
\left(\frac{p^2}{2m_j}+U^{(j)}({\bf r})-
\mu^{(j)}\right)\right]+1\right\}^{-1},
\label{equilibrium}
\end{equation}
where $\beta=1/k_BT$ and $\mu^{(j)}$ is
the chemical potential of species $j$~\cite{madda}.
The particle densities entering $U^{(j)}({\bf r})$
are to be determined self-consistently by
integration over momenta, and the momentum
distributions of the two clouds need generating.

To exploit the axial symmetry of the system, 
we first define the angularly integrated particle densities 
${\cal N}^{(j)}(r,z)=2\pi r n^{(j)}(r,z)$
on a $\{r,z\}$ 
grid in cylindrical
coordinates and we move to a particles-in-cell
description by locating a number
${\cal N}^{(j)}(r,z)\Delta r\Delta z$ of fermions
in each cell of volume $\Delta r\Delta z$~\cite{marilu}.
A momentum distribution with
low statistical noise is generated by representing
each fermion by means of $N_q$ computational particles (``quarks'').
The $i$-th quark is located at point
$\{p_{ir},p_{i\theta},p_{iz}\}$ in momentum space by
using a Monte Carlo sampling and making sure that
each cell of volume $h^3$ is occupied by no more than $N_q$ quarks.
This control in 3D phase space is 
transferred to 2D by imposing a maximum number
$wN_q$ of quarks in the 2D cell $\{\Delta r,\Delta z,
\Delta p_r,\Delta p_z\}$ of volume $h^2$. 
Here the weight $w={\rm int}(2p_Fr/\hbar)$ takes into account
that the number of available cells in 2D
depends on the radial position and on the number of particles through 
the Fermi momentum $p_F= \sqrt {2m\hbar k_BT_F}$.

In the propagation step the two clouds are rigidly displaced from the center
of the trap along the $z$ direction and start evolving in time
in the $\{r,z\}$ plane by performing oscillations at their respective
frequencies.
We exploit  
the fact that by symmetry the average value
of $p_{\theta}$ for each value of $r$ is zero and does not change in time,
to decouple the variable $p_{\theta}$ from the equations of motion. 
We make the approximation that the angular momentum of each quark
is left unchanged by the collisional events and take into account
the third dimension
in the exclusion principle by suppressing collisions whose
final states are occupied by more than $wN_q$ quarks.
At each time step the quarks are moved by the 
confinement and the mean-field 
forces by using a Verlet algorithm
on a grid of mesh spacing $dx>v\,dt$,
where $v$ is a typical
particle velocity and $dt$ the time step.
This inequality is dictated by accuracy and stability
of the propagation step~\cite{marilu}. 
No exclusion constraint is applied during this step:
this does not lead to any appreciable deviation from 
Fermi statistics down to 0.2 $T_F$, where 
the system is still sufficiently dilute.
This has been checked by  
monitoring violations of the Pauli principle at each time step.

Finally we come to the collision step, which involves most of the innovative
aspects of our scheme.
Collisions are tracked on a grid of mesh spacing $\delta x$ of the order
of the de Broglie wavelength $\lambda_B$, 
which is smaller than the particle mean free path $l$
and larger than the spacing $dx$ of the propagation mesh
($l>\delta x\sim\lambda_B>dx$).
The first inequality enhances the statistical accuracy 
of the collision step, whereas coarse graining with respect to the
propagation step avoids the need for
Pauli-principle constraints, at least down to 0.2 $T_F$.

Before turning to the results we add 
some technical details on how a speed-up of the code
by several orders of magnitude has been achieved. 
The number of probable collisions between all possible pairs of 
quarks belonging to the two species in each cell of volume $h^2$ 
is evaluated at each computational time step as
$dN_{coll}= dt \sqrt {N_1 N_2} \sum_{ij} v_{ij} \sigma_{ij}$.
Here $v_{ij}$ is the magnitude of the relative speed of particle $i$ of species
$1$ and particle $j$ of species $2$, $\sigma_{ij}$ is the corresponding 
differential
cross section, and $N_1$ ($N_2$) is the number of particles of species 
$1$ ($2$) in the given spatial cell.
If $dN_{coll}<1$, the collision probability is accumulated 
over the subsequent time steps until an integer number 
$dI_{coll} \equiv {\rm int}(dN_{coll})$ 
of collisions occurs.
Within each spatial cell the pairs of particles are 
ordered according to the value 
of their classical collision probability, from largest to smallest.
The acceptance rate
of the MonteCarlo sampling is enhanced by two-three orders of magnitude 
at each step by filtering out pairs with classical probability 
smaller than a given threshold. This is  
a form of importance sampling and the 
threshold is dynamically adjusted cell-by-cell in such
a way as to guarantee the correct supply of $dI_{coll}$ 
collisions at each time step.
The collision probability becomes smaller than the classical one
after multiplication by the quantum suppression factor 
$1-N(r,z,p_r,p_z)/(wN_q)$
due to the occupancy of the final state, 
and consequently the selection of the most likely 
pairs to collide becomes essential. 
To this purpose each particle is allowed to collide only with the 
partner which maximizes the product $v_{ij}\sigma_{ij}$. 
The original pool of $N_1N_2$  collisions is cut down to $N_1$ collisions only, with a 
resulting additional speed-up of at least one order of magnitude.

\paragraph*{Results.}
As a first application of the numerical method we have
considered 200 magnetically trapped $^{40}$K atoms, 
which are represented by a total
number of 4$\times10^3$ quarks. 
As in the JILA experiments~\cite{gensemer}, the atoms are equally shared
among two different spin states ($m_f=9/2$ and $m_f=7/2$)
in harmonic traps with slightly different
longitudinal frequencies 
($\omega_{9/2}=2\pi\!\times\!19.8\,\,$s$^{-1}$ and
$\omega_{7/2}=2\pi\!\times\!17.46\,\,$s$^{-1})$.
When the two clouds after initialization are 
rigidly displaced from the center
of the traps,
in the absence of interactions they would keep oscillating at their 
respective trap frequencies without damping.

\begin{figure}

\centering{
\subfigure[]{\epsfig{file=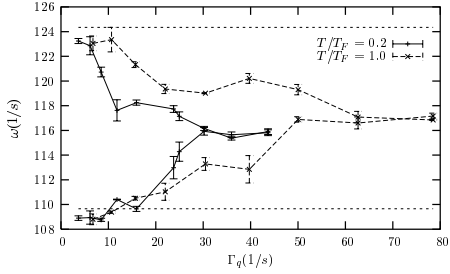,width=0.7\linewidth}}
\subfigure[]{\epsfig{file=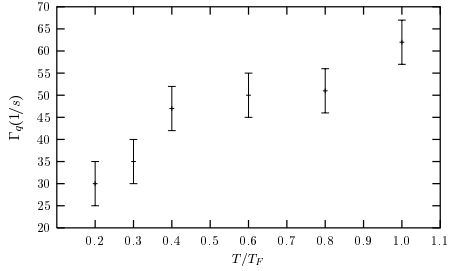,width=0.7\linewidth}}}
\caption{(a) The  oscillation frequencies $\omega$ (in units of s$^{-1}$) 
as functions of the quantum 
collision rate $\Gamma_q$ (in units of s$^{-1}$) for 
the two components of the gas at $T=T_F$ ($\times$) and 
$T=0.2\,T_F$ ($+$). The horizontal dashed lines show the bare 
trap frequencies.
(b) The collision rate $\Gamma_q$ at frequency locking
as a function of temperature $T$ (in units of $T_F$).}
\label{fig1}
\end{figure}

\begin{figure}
\centering{
\epsfig{file=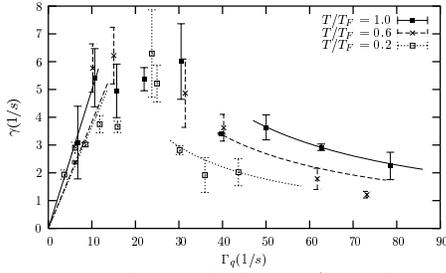,width=0.7\linewidth}}
\caption{The damping coefficient $\gamma$ (in units of s$^{-1}$) as a function
of the collision rate $\Gamma_q$ (in units of s$^{-1}$) for $T=T_F$ 
(filled squares), $T=0.6T_F$ (crosses) and $T=0.2T_F$ (empty squares).
In the collisionless region $\gamma$ has been fitted by the function
$\gamma(\Gamma_q)=\alpha_1\Gamma_q$ and in the collisional regime
by the function $\gamma(\Gamma_q)=\alpha_2/\Gamma_q$:
the fits are shown by a
continuous line for $T=T_F$, a dashed line for $T=0.6T_F$ and a dotted line
for $T=0.2T_F$.} 
\label{fig2}
\end{figure}

The collision rate $\Gamma_q$ is independently varied by changing the
 scattering
length $a$, thus mimicking the exploitation of a suitable Feshbach 
resonance~\cite{fesh}.
To offset the difference between the number of atoms 
used in the simulation ($N=200$) and that
in the experiments ($N_{exp}\sim10^6$), the reference value of the
off-resonant scattering length is scaled by a factor $(N_{exp}/N)^{1/2}
\sim 10^2$, thus producing a system
with essentially the same collision rate and mean field potential
as in Ref~\onlinecite{gensemer}.
The transition from the collisionless to the collisional regime,
as driven by varying the scattering length at various temperatures, is 
shown both in the plot of the frequency of the dipole 
mode for the two components in Fig.~\ref{fig1}(a) and in the plot of 
the damping rate $\gamma= 1/\tau$ of the axial motion of 
the centers of mass $z_{\rm cm}^{(j)}(t)$ in Fig.~\ref{fig2}.

In Fig.~\ref{fig1}(a) the oscillation frequencies $\omega_j$ at  
$T=T_F$ and $T=0.2\,T_F$, with $k_BT_F=\hbar\omega_{9/2}(6N_{9/2})^{1/3}$, 
are evaluated by fitting  $z_{\rm cm}^{(j)}(t)$ with the functions 
 $z_0\cos(\omega_j t)e^{-\gamma t}$.
At very low $\Gamma_q$ the dipole mode frequencies are given by 
the corresponding trap frequencies with 
a shift due to the mean-field potential.
At intermediate values of $\Gamma_q$ 
the data points
show large fluctuations, due to the fact that in 
this region just a few collisions can drastically alter  
the motion of the clouds.
However, the trend towards a locking of the two dipole
mode frequencies at large $\Gamma_q$ is very clear and 
the location of the locking is identified
with reasonable accuracy.

The main new physical result of this study is the
shift of the locking transition to lower $\Gamma_q$  
as temperature is decreased. This is shown in Fig.~\ref{fig1}(b).
This effect is related to Fermi statistics: at low temperature 
the collisions involve particles on a narrower region
around the Fermi level and
have a greater impact on the global dynamics of the gas. 
A smaller number of collisions is needed
to produce locking of the two interacting species.

The damping rate $\gamma$ obtained from the fit of
$z^{(j)}_{cm}(t)$ is essentially the same for the two components
and is shown in Fig.\ref{fig2}.
We have also evaluated the correlation function
$\phi(t)=<|{\cal Z}_{\rm cm}^{(j)}(t')||{\cal Z}_{\rm cm}^{(j)}(t'+t)|>$
between the magnitude of the turning points ${\cal Z}_{\rm cm}^{(j)}$,
which decays exponentially as $\exp{(-\gamma t)}$
at large $t$.
This estimate of $\gamma$ is in good agreement with 
that obtained from the center-of-mass motions,
but yields a less noisy signal in the intermediate region.
In the collisionless regime 
the damping rate increases linearly with $\Gamma_q$, while in the
collisional one it scales like $\Gamma_q^{-1}$. This is seen 
in Fig.~\ref{fig2}, which also shows again that
the hydrodynamic regime is reached at lower $\Gamma_q$
as $T$ is lowered.

\begin{figure}
\centering{
\epsfig{file=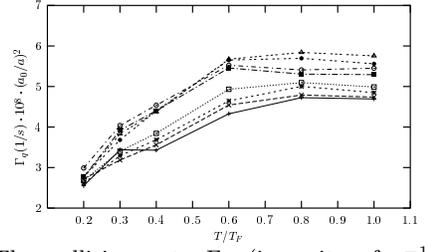,width=0.7\linewidth}}
\caption{The collision rate $\Gamma_q$ (in units of s$^{-1}$) scaled 
by the factor 
$10^{8} (a_0/a)^2$ with $a_0$ being the Bohr radius, as a function
of temperature $T$ (in units of $T_F$) for various 
values of the scattering length $a$.
From bottom to top: 
$a=(12,15,18,21,24,27,30,33)\times 10^3a_0$.}
\label{fig3}
\end{figure}

The collision rate $\Gamma_q$ 
can be driven by cooling at fixed scattering length,
as is seen from Fig.~\ref{fig3}. 
The various curves, after scaling by a factor proportional
to the
classical cross-section, show a residual weak dependence on $a$,
which is due to the mean-field interaction 
between the two overlapping fermionic components.
The trend of these curves is mostly classical and only at very low
temperature ($T=\,$0.3-0.2 $T_F$) Pauli blocking becomes manifest, as signalled
by the collapse of the curves into a single one.
Our results thus suggest new  experiments,
in which the collision rate is changed 
{\it via} a thermal drive. 

The implication of the results shown in Fig.~\ref{fig2} and 
Fig.~\ref{fig3} is that this thermal drive
rests upon the increasing
importance of a decreasing set of ``strategic'' collisions
involving particles in states clustered around the Fermi level.
Our data also indicate that thermal
cooling  will need to reach temperatures below  $0.2 \; T_F$
in order to see the complete 
transition from the collisional to the collisionless regime.
Efforts to develop a concurrent code, including the Pauli principle
in the Lagrangian evolution to treat the collisional Fermi gas well 
below $0.2\,T_F$, are currently under way.

\noindent
{\bf Acknowledgement:}
We thank Dr. A. Minguzzi for helpful discussions
and acknowledge support from INFM through the PRA2001 Program.

\end{document}